\documentclass[aps,pre,twocolumn,showkeys,showpacs,superscriptaddress]{revtex4-1}

\usepackage[nottoc,numbib]{tocbibind}
\usepackage[T1]{fontenc}
\usepackage[utf8]{inputenc}
\usepackage{amssymb}
\usepackage{pifont}
\usepackage{textcomp}
\usepackage[centertags]{amsmath}
\usepackage{graphicx}
\graphicspath{{figures/}{logos/}}
\DeclareGraphicsExtensions{.pdf,.png,.jpg}
\usepackage[outdir=./]{epstopdf}
\usepackage{booktabs}
\usepackage{url}
\usepackage{epsfig}
\usepackage{textcomp}
\usepackage{subfigure}
\usepackage{lmodern}
\usepackage[squaren]{SIunits}
\usepackage{placeins}
\usepackage{color}
\usepackage{listings}
\usepackage{floatpag}
\usepackage{needspace}
\usepackage[]{natbib}
\usepackage{verbatim}
\usepackage{array}
\usepackage{hyperref}

\newcommand{\bcdot}{\hspace{-3.5pt}\ensuremath{\cdot}\hspace{-3.5pt}}
\newcommand{\pderDbl}[3]{\ensuremath\frac{\partial^2 #1}{\partial #2\partial #3}}
\newcommand{\pder}[2]{\ensuremath\frac{\partial #1}{\partial #2}}

\newcommand{\Rot}[1]{\ensuremath{ \nabla \times #1 }}
\newcommand{\Div}[1]{\ensuremath{ \nabla \cdot #1 }}

\newcommand{\abs}[1]{\left|#1\right|}

\newcommand{\kl}[1]{\ensuremath{\left(#1\right)}}
\newcommand{\ekl}[1]{\ensuremath{\left[#1\right]}}

\makeatletter
\newcommand{\thickhline}{%
    \noalign {\ifnum 0=`}\fi \hrule height 1pt
    \futurelet \reserved@a \@xhline
}
\newcolumntype{"}{@{\hskip\tabcolsep\vrule width 1pt\hskip\tabcolsep}}
\makeatother

\begin{document}


\title{Magnetic Reconnection in Turbulent Diluted Plasmas}


\author{N. Offeddu} 
\email{nico.offeddu@gmail.com} 
\affiliation{ ETH
  Z\"urich, Computational Physics for Engineering Materials, Institute
  for Building Materials, Wolfgang-Pauli-Strasse 27, HIT, CH-8093 Z\"urich
  (Switzerland)}

\author{M. Mendoza} 
\email{mmendoza@ethz.ch} 
\affiliation{ ETH
  Z\"urich, Computational Physics for Engineering Materials, Institute
  for Building Materials, Wolfgang-Pauli-Strasse 27, HIT, CH-8093 Z\"urich
  (Switzerland)}

\date{\today}

\begin{abstract}
We study magnetic reconnection events in a turbulent plasma within the two-fluid theory. By identifying the diffusive regions, we measure the reconnection rates as function of the conductivity and current sheet thickness. We have found that the reconnection rate scales as the squared of the inverse of the current sheet's thickness and is independent of the aspect ratio of the diffusive region, in contrast to other analytical, e.g. the Sweet-Parker and Petscheck, and numerical models. Furthermore, while the reconnection rates are also proportional to the square inverse of the conductivity, the aspect ratios of the diffusive regions, which exhibit values in the range of $0.1-0.9$, are not correlated to the latter. Our findings suggest a new expression for the magnetic reconnection rate, which, after experimental verification, can provide a further understanding of the magnetic reconnection process. 
\end{abstract}


\maketitle  

%
\section{Introduction}

During magnetic reconnection processes magnetic field lines in a plasma  recombine, and thereby abruptly change their topology. Such processes are present in a wide range of physical systems in very different regimes, including, among others, fusion reactors \citep{MR_in_tokamak}, Earth's magnetic belt, solar corona and chromosphere \citep{gonzalez2016magnetic}. Magnetic reconnection is thought to be the underlying cause of solar coronal mass ejection and polar auroras \citep{priest}. Due to a very efficient conversion of magnetic energy into kinetic energy, the time and energy scales involved can be of an extreme magnitude \citep{energy}. Since the early 50s, this phenomenon has been investigated under different perspectives, both in theoretical and experimental studies. Although great progress in computational, experimental and observational physics has been made \citep{yamada_rev_1,yamada_rev_2}, many crucial aspects of magnetic reconnection are not yet fully understood, e.g., the interplay of small-scale physics in the diffusive region with the global dynamics of the system \citep{yamada_two_fluid}, or the role played by turbulence and three-dimensional asymmetries \citep{perspectives,3D}. 

In early attempts to explain this phenomenon, Sweet \citep{sweet} and Parker \citep{parker} set up the frame upon which modern studies of magnetic reconnection are  based.  They estimated the dependence of the reconnection rates on geometrical quantities of the process, and described the first simple two-dimensional steady-state scenarios. In particular, they contemplated two opposed magnetic flux tubes of length   $L$ pushed together to a distance $\delta$, thus creating a central diffusive region with vanishing magnetic field. As it can be seen in Ref. \citep{parker}, a geometry with small aspect ratio, i.e. $\delta \ll  L$, is assumed. Parker showed that the plasma is drawn into the diffusive region across the magnetic flux lines and expelled at much higher velocity from the sides, together with recombined magnetic lines. He expressed the reconnection rate as the ratio of the inflow velocity to the outflow velocity $v_{\text{in}}/v_{\text{out}}$.  Later, Petscheck extended this work by introducing stationary slow mode shocks connecting the in- and outflow regions \cite{petscheck}. Since the plasma flow is not bound to flow only across the magnetic field lines of the flux tubes, the diffusive region has a  larger aspect ratio, and an increased reconnection rate. Nevertheless, both the Sweet-Parker and Petscheck reconnection rates fail to recover higher rates observed experimentally. The Sweet-Parker and Petscheck approaches, their limits, and their relevance to this paper are briefly outlined in Sec. \ref{sp_model}. 


One of the aspects not considered by the Sweet-Parker and Petscheck approaches, is the influence of turbulence. Especially in astrophysical plasmas, turbulence is common, as, e.g., in situ observations of Earth's bow shock  showed \citep{modern_turbulence}. Recently, many numerical studies tend to focus on this aspect, e.g. in  simulations of Earth's magnetosphere \citep{sim_earth_magnetosph}. Earth's magnetosphere is also the subject of study of a mission by NASA, the Magnetospheric Multiscale Mission (MMS) \citep{burch_magnetospheric}. MMS consists of four satellites currently flying in formation through the day-side magnetopause and the magnetotail. For the first time, the diffusive region in Earth's magnetosphere during reconnection processes is being measured with enough space and time resolution to appreciate small scale physics. 
Turbulence in the diffusive region, is shown to play a role in the enhancement of magnetic reconnection, e.g. in 2D simulations of single reconnection sites with a background turbulence \citep{Loureiro}. Some numerical studies also focus on dynamical systems with multiple reconnection sites rather than considering a single reconnection event. For instance, in Ref.~\citep{turb_2d}, it is shown that reconnection rates are enhanced by global turbulence. However, the mentioned 2D model were based on magnetohydrodynamics (MHD), neglecting relevant processes coming from the interaction between electrons and ions, e.g. the Hall effect. 

Recent experimental and theoretical studies on two-fluid plasmas, pointed out how the interaction of electrons and ions through Coulomb collisions is decisive for the dynamics of the system. In the case of low density plasma, such as astrophysical plasmas in Earth's magnetosphere, as studied by MMS \citep{burch_magnetospheric}, the one-fluid MHD approach fails to recover essential dynamics and electron-ion interactions \citep{perspectives}. For this, the two-fluid approach is more appropriate  as it presents several advantages in comparison to other methods, e.g., the MHD assumption of a vanishing electrical field within the plasma is relaxed \citep{yamada_two_fluid}. This is why the two-fluid model is very well suited for the study of plasmas in which the differential flow of electrons and ions plays a major role. For example, during magnetic reconnection events in highly diluted plasmas with low collisionality between electrons and ions, the different species of particles are allowed to follow different paths. The differential flow generates a Hall effect that affects the dynamics in the diffusive regions in a non-linear manner \citep{nostro_1}. In the MRX experiment \citep{yamada_two_fluid}, it has been shown by tuning the density of the plasma that this effect directly influences the shape of the diffusive region and the reconnection rate. By increasing the density of the plasma, and thus the collision frequency between electrons and ions, in the mentioned work the diffusive region recovers a Sweet-Parker-like profile with a small aspect ratio. When the plasma is sufficiently diluted, i.e., a low collision frequency, the diffusive region assumes a Petschek profile with a larger aspect ratio and an increased reconnection rate. A more detailed discussion on the two-fluid model and its relevance to this work follows in Sec. \ref{sec:2_fluid}. 

The goal of our work is to study multiple magnetic reconnection processes in a turbulent plasma, by identifying the diffusive regions and propose a relation for the magnetic reconnection rate as function of the plasma parameters, within a more microscopic description using two-fluid plasma theory. Our numerical simulations are implemented using a Lattice-Boltzmann code, as described in Ref. \citep{nostro_1}, which has been proven to be an effective and flexible method for the study of plasma dynamics \citep{lb_mhd_vecchio,miller_2}. In our work, various magnetic reconnection processes are induced by turbulence in a system which is larger than the typical flux tube length $L$. We measure the reconnection rates in the diffusive regions identified by an algorithm. A strong correlation between the reconnection rates and the inverse squared conductivity, as well as between the reconnection rates and the inverse of the squared current sheet's thickness can be noted. This findings differ from Sweet-Parker's and Petscheck's models. Another interesting fact, is that while changing the conductivity of the plasma over an order of magnitude, the aspect ratio did not present any noticeable change.

Our work is organized as follows: In Sec. \ref{sp_model} we review the theory on the first models on magnetic reconnection. In Sec. \ref{sec:2_fluid}, we describe the adopted mathematical method, and how this has been implemented in our numerical simulations. The results are then analyzed in Sec. \ref{sec:simulations}. Finally, in Sec. \ref{sec:concl},  we discuss our results and draw some conclusions.

\subsection{Sweet-Parker and Petscheck Models}
\label{sp_model}
In MHD, plasma is considered  as a fluid composed of charge carriers such as electrons and ions, but with no net charge. The main quantities of the plasma flow and the electromagnetic fields are often given in dependence of the magnetic field. This is viewed as the primary field from which all the others are computed. 

Let us start with the Maxwell equations,
\begin{align}
  \begin{aligned}
    \Rot{\vec{E}} &=-\pder{\vec{B}}{t} \quad ,&
    \Rot{\vec{B}} &=\mu\vec{j} + \mu\epsilon\pder{\vec{E}}{t} \quad ,&
    \\
    \Div{\vec{E}} &=\frac{1}{\epsilon}\rho_{\varepsilon} \quad ,&
    \Div{\vec{B}} &=0 \quad ,
  \end{aligned}
  \label{eq:maxwell}
\end{align}
where $\vec{E}$ and $\vec{B}$ are the electric and magnetic field, $\vec{j}$ and $\rho_{\varepsilon}$ the current and charge densities, and $\mu$ and $\epsilon$ the magnetic permeability and electric permittivity, respectively. In the non-relativistic regime, i.e., the fluid's motion is confined to velocities much smaller than the speed of light, the electrical currents in the plasma can be expressed with the help of the inertial-frame Ohm's law, which reads
\begin{equation}
	\vec{j} = \sigma\kl{\vec{E}+\vec{v}\times\vec{B}} \quad , 
	\label{eq:inertial_ohm}
\end{equation}
or from Ampere's law, by taking the curl of the magnetic field and neglecting the partial time derivative:
\begin{equation}
	\vec{j} = \frac{1}{\mu} \Rot{\vec{B}} \quad .
\label{eq:curr_2}
\end{equation}
By combining equations \eqref{eq:curr_2} and \eqref{eq:inertial_ohm}, we obtain the electric field
\begin{equation}
	\vec{E} = - \vec{v}\times \vec{B} + \frac{1}{\sigma\mu} \Rot{\vec{B}} 
	\label{eq:el_inertial_field}
\end{equation}
and, by taking the divergence from the latter, the charge density 
\begin{equation}
	\rho_{\varepsilon} = -\epsilon \Div{\kl{\vec{v}\times\vec{B}}} 
\end{equation}
can be expressed in terms of the magnetic field. The time evolution of the magnetic field closes this set of equations, which can all be determined for a known initial configuration of $\vec{B}$. Taking the curl of Eq.~\eqref{eq:el_inertial_field}, using the Maxwell equations for the rotation of $\vec{E}$ and the divergence of $\vec{B}$, yields the \emph{induction equation}:
\begin{equation}
	\pder{\vec{B}}{t} = 
	\Rot{\kl{\vec{v}\times\vec{B}}} + \eta_m \nabla^2 \vec{B} \quad ,
	\label{eq:induction_eq}
\end{equation}
where $ \eta_m \equiv 1/\kl{\mu\sigma} $, known as magnetic diffusivity, is proportional to the inverse of the Spitzer conductivity $\sigma$ and the permeability of free space $\mu$. The electromagnetic  fields, as well as $\vec{j}$ and $\rho_\varepsilon$, can now be calculated starting with the magnetic field $\vec{B}$.

From the induction equation, it can be seen that when the magnetic diffusivity is low, the magnetic field is mainly advected by the flow of the plasma. This is the case when the magnetic Reynolds number $\text{Re}_m = \frac{V L}{\eta_m}$, where $L$ and $V$ are the characteristic size and velocity of the system, is much greater than one. In this regime, the magnetic field is tied to the plasma which drags its lines along with the flow. Due to this effect, the magnetic field lines are not able to diffuse and rearrange, which is also known as \emph{Alfv\'en's frozen in theorem} \citep{frozen_in}. This is only possible if the local conditions are far from the assumed ones. In the late 1940s, Sweet's interpretation  of the phenomenon \citep{sweet} was that when two magnetic flux tubes are pushed together by external forces, the strong curvature in the magnetic field makes the last term on the right hand side of Eq. \eqref{eq:induction_eq} non-negligible. The magnetic field lines are then able to diffuse in a small region (the diffusive region)  between the flux tubes, and recombine their topology. This generates a current sheet within the plasma, and once the recombined magnetic lines leave the diffusive region, material is accelerated to high velocities by the strong curvature in the magnetic field. A characteristic quantity of the system is the reconnection rate $\Gamma$, which is the ratio between the inflow and outflow velocity of the plasma. In the early 1950s, Parker  made the following estimation \citep{parker} of the reconnection rate in the setup proposed by Sweet. Consider two flux tubes in the $X-Y$ plane, of  length of $L$, being pushed to a distance $\delta$. An $\vec{E}\times\vec{B}$ drift will generate a flow of material towards the center of the system. In the limit of incompressibility, the amount of plasma flowing in will match the outflowing plasma, thus the approximation $L  \abs{v_{\text{in}}} = \delta  \abs{v_{\text{out}}}$ holds throughout the process. This means that the Sweet-Parker reconnection rate can be expressed as
\begin{equation}
\Gamma_{\text{SP}} = \frac{\delta}{L}
\label{eq:gamma_from_v_ratio}
\end{equation}
If $L>>\delta$, i.e., if the two flux tubes are pushed closely together, it follows that the outgoing velocity is much greater than the inflow velocity.

The reconnection rate, can also be expressed in terms of characteristic plasma parameters in the following way. According to Eq. \eqref{eq:curr_2}, the curl of the magnetic field will generate an out of plane current $J_z=\vec{\nabla}\times \vec{B}/\mu\propto B_0/ \mu \delta$, where $B_0$ is the characteristic field strength. With a negligible magnetic field in the diffusive region (which extends across a depth of $\delta$), Ohm's law simplifies to $\vec{J}=\sigma\vec{E}$, with  the conductivity $\sigma$. As already mentioned, at the border of the diffusive region, an $\vec{E}\times\vec{B}$ drift will generate a flow of material towards the center of the system, across the magnetic field lines which are, in that region, negligible.  Using Ohm's law and the curl of the magnetic field, the velocity of inflowing material can be estimated as
\begin{equation}
\abs{v_{\text{in}}}\propto \vec{E}\times\vec{B}/\vec{B}^2 \propto\frac{1}{\mu \delta \sigma} \quad .
\end{equation} 
Neglecting the relatively small inflow velocity in the diffusive region and balancing the magnetic and dynamic pressure inside and outside of the region, 
\begin{equation}
\frac{B_0^2}{2\mu} \propto \frac{1}{2}\rho |\vec{V}_{\text{out}}|^2 \quad ,
\end{equation}
yields an outflow velocity of
\begin{equation}
\abs{\vec{v}_{	\text{out}}} \propto \frac{B_0}{\sqrt{\rho \mu}} \equiv V_\text{A} \quad ,
\end{equation}
where $V_\text{A}$ is the \emph{Alfv\'en velocity} of the system. By taking the ratio of inflow and outflow velocity, the Sweet-Parker reconnection rate is calculated as
\begin{equation}
\Gamma_{\text{SP}} \propto\abs{\frac{\vec{v}_{\text{in}}}{\vec{v}_{\text{out}}}} \propto \frac{1}{\mu V_{\text{A}} \delta\sigma} \quad .
\label{eq:gamma_from_fields}
\end{equation}
Note that multiplying the reconnection rate from Eq. \eqref{eq:gamma_from_v_ratio} with the one obtained in Eq. \eqref{eq:gamma_from_fields}, one has
\begin{equation}
\Gamma_{\text{SP}}  \propto \kl{\mu V_{\text{A}} L \sigma}^{-\frac{1}{2}}  \equiv \frac{1}{\sqrt{S}}
\quad ,
\label{eq:sw_rate}
\end{equation}
where $S$ is known as the  \emph{Lundquist number} of the system. 

A particular feature of the Sweet-Parker reconnection is that the diffusive region presents a very thin depth $\delta$ compared to its elongation $L$.  Some years later, Petschek proposed a model \citep{petscheck} based on slow-mode shocks, in which plasma is not bound to diffuse only in the central region of the system. This opens the outflow of plasma, thus increasing the reconnection rate, which in his model is calculated as
\begin{equation}
\Gamma_{\text{P}} \propto \frac{1}{\ln{S}} \quad .
\label{eq:petsch_rate}
\end{equation}
In resistive MHD, numerical studies only recover the Petscheck aspect ratio when in the diffusive region a locally larger resistivity is used. In experimental studies, e.g. \citep{yamada_two_fluid}, it is shown that the Petscheck profile appears, although without evident signs of slow shock modes, when the collision between ions and electrons is reduced, and a differential flow is allowed among electrons and ions.

There is a general consensus \citep{perspectives} that simple models such as two-dimensional setups or the MHD approximation are insufficient to quantitatively describe empirical observations of reconnection phenomenona. Turbulence \citep{turbulence_1}, the generalized Ohm's law \citep{yamada_two_fluid} and setups with strong three-dimensional and asymmetrical character \citep{3D} are thought to be among the underlying causes of the discordance between the observed reconnection rates and theoretical estimations. As already mentioned, an example for one of the important effects which is neglected by MHD, is the differential flow of species composing the plasma, or the Hall effect thereby generated \citep{yamada_two_fluid}.

Magnetic reconnection is also observed in exotic and extreme environments \citep{perspectives}, such as gamma-ray bursts. Relativistic effects, radiative pressures, Compton radiation and interactions between elementary particles all have to be accounted for \citep{uzdensky}. This work is restrained to non-relativistic environments, as in the the solar corona or Earth's magnetosphere. 

\subsection{Two-fluid generalization}
\label{sec:2_fluid}
MHD relies on the assumption that the electric fields within the plasma are canceled out almost instantaneously by currents in the neutral plasma. This is approximation holds for a conductivity which tends to infinity, due to which the charges in the plasma can be displaced very quickly as soon as an internal field is generated. However, if the temperature of the plasma is high enough or the plasma is highly diluted, the conductivity drops (see Eq. \eqref{eq:conductivity_def})  and the electron and ion momenta must be taken into account separately. In such a case, the approach is to consider the plasma as composed by two (or more) species. 

The generalization for a fluid composed by two species of particles starts by considering the conservation of momentum for the particles' species,
\begin{align}
\label{eq:gen_mom_eq}
m_s n_s \kl{\pder{\vec{v}_s}{t} + \kl{\vec{v}_s\bcdot\nabla}\vec{v}_s} &=  
n_s q_s\kl{\vec{E} + \vec{v}_s\times\vec{B}} \nonumber \\
&- \nabla P_s + n_s m_s \eta_s \nabla^2 \vec{v}_s \nonumber \\
& -\nu\rho_0\kl{\vec{v}_s - \vec{v}_{\bar{s}}}  \nonumber \\
& \vec{F}^{\rm ext.}_s \quad ,
\end{align}
where $m_s$ and $n_s$ are the particles' mass and density, $\vec{v}_s$ the velocity, $P_s$ the pressure, $q_s$ the electric charge, and $\eta_s$ the kinematic viscosity. Here, the indices $s=0,1$ and  $\bar s\equiv \kl{s+1}\text{mod}2$ indicate the electrons ($s=0$) and  ions ($s=1$). Note that the exchange of momentum density between the populations via Coulomb collisions is proportional to the velocity difference and the collision frequency $\nu$, the third term on the right hand side of Eq.~\eqref{eq:gen_mom_eq}. Since the transfer is symmetric, the same expression is valid for both  populations of the plasma. External forces can be considered by adding an extra forcing term, $\vec{F}^{\rm ext.}_s$.

Combining the momentum and continuity equations for the two populations, the generalized Ohm's law can be derived as \citep{Somov2007}: 
\begin{align}
\vec{E} + \vec{v}\times\vec{B} 
&= - \frac{m_0 m_1}{q_0 q_1} \pder{\,}{t} \kl{\frac{\vec{j}}{\rho_m}} &\kl{\text{{electron inertia}}  }\nonumber \\
& + \frac{1}{\rho_m} \kl{\frac{m_1}{q_1}+\frac{m_0}{q_0}}\vec{j}\times\vec{B}   &\kl{\text{{Hall term}}  } \nonumber \\
& - \frac{m_0 m_1}{\rho_m q_0 q_1}\kl{\frac{q_1}{m_1}\nabla P_1 + \frac{q_0}{m_0}\nabla P_0 }  \hspace{-0.7cm} &\kl{P_1\text{{ gradient}}  } \nonumber \\
& - \kl{1 - \frac{q_0 m_0}{ q_1 m_1}} \frac{m_0 m_1}{\rho_m q_0 q_1} \nu \, \vec{j}\,\quad .   &\kl{\text{{Ohm's term}}  }  \nonumber \\
\label{eq:general_ohm}
\end{align}

In most cases, the equation can be simplified considerably. For example, since $\frac{q_1}{m_1}<<\frac{q_0}{m_0}$, the pressure term of the ionic population is generally much smaller than the pressure gradient for electrons and is therefore often neglected. In a steady state, in the limit of quasi-neutrality of the plasma and in absence of magnetic fields, Eq. \eqref{eq:general_ohm} reduces to
\begin{equation}
\vec{E} = \underbrace{ \frac{m_0 \nu}{q_0^2 n_0 }  }_{\text{resistivity}} \vec{j}
        = \frac{1}{\sigma}\,\,\vec{j} \quad ,
	\label{eq:conductivity_def}
\end{equation}
where $\sigma$ is the conductivity of the plasma.
It follows \citep{fridman} that in an ionized plasma, in which the electrons and ions collide faster than the typical gyro-frequencies, the conductivity can be regarded as a scalar defined as in  Eq. \eqref{eq:conductivity_def}. 

Eqs. \eqref{eq:maxwell}, \eqref{eq:gen_mom_eq}, \eqref{eq:general_ohm}, and the equation of state $P_s = \bar{P}_s (\rho_s/\bar{\rho}_s)^\gamma$, with $\gamma$ the polytropic index and $\bar{\rho}_s$ and $\bar{P}_s$ some characteristic density and pressure, respectively, represent the set of equations of the two-fluid theory. They are recovered by the numerical model used in this work \citep{nostro_1} which is shown \citep{miller_1,miller_2} to correctly reproduce the Maxwell equations \eqref{eq:maxwell}, the incompressible and viscous fluid equations \eqref{eq:gen_mom_eq} and the generalized Ohm's law \eqref{eq:general_ohm} with a kinematic viscosity 
\begin{equation}
\eta_s = c_s^2 \kl{\tau_s - \frac{1}{2}} \Delta t = \frac{1}{3} \kl{\tau_s - \frac{1}{2}} \frac{\Delta x^2}{\Delta t}\quad ,
\label{eq:kin_visc}
\end{equation}
with a speed of sound $c_{th}=(1/\sqrt{3}) \Delta x/\Delta t$ and $\tau_s$ a relaxation parameter to tune the kinematic viscosity. In this work, we use $\gamma = 1$, and $P_s = \rho_s c_{th}^2$.  

A major difference to the MHD approach, is that in a two-fluid model electrons and ions are allowed to follow different paths in their flow towards and out of the diffusive region. Numerical and experimental studies, show that this is indeed the case during magnetic reconnection. For example, in \citet{yamada_two_fluid} it is described how the motion of the ions is characterized by a higher inertia, thus  demagnetizing sooner than the electrons while approaching the diffusive region. On the contrary, electrons follow the magnetic lines deeper into the diffusive region. This creates a differential flow that generates a quadrupole magnetic field. In a two-fluid plasma, it is shown \citep{nostro_1} that the Hall effect can differ substantially from the one in a classical conductor, and the quadrupole field affects the dynamics in the diffusive region in a non-linear manner, thereby increasing the reconnection rate.

\section{\label{sec:simulations}Numerical simulations}
\begin{figure}
		\centering
		\includegraphics[width=0.45\textwidth]{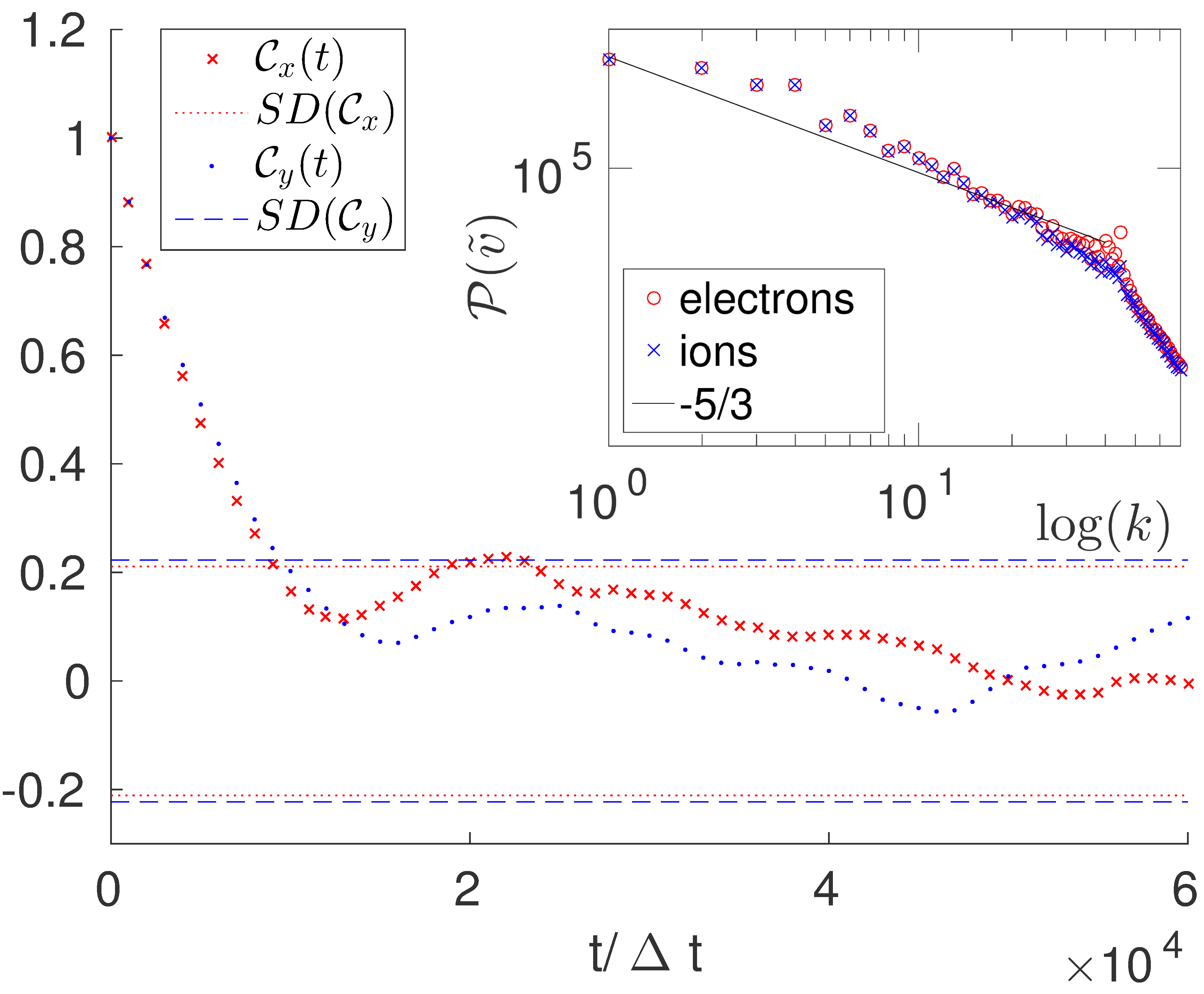}
		\caption{Decay of the correlation functions $C_x\kl{t}$ and $C_y\kl{t}$ defined in Eq. \eqref{eq:corr_fun}. The dotted lines represent the standard deviations of the two correlation functions. In the upper-right panel, the power spectrum of the ionic and electronic energy is shown. We observe that in both cases the characteristic inverse energy cascade has a slope of approximately $-5/3$, as described in \citet{boffetta}.
}
	\label{fig:correlation}
\end{figure}
It is important to mention that we consider a turbulent plasma with multiple reconnection events, instead of a single reconnection site at steady state. This means that different reconnection processes can happen at different locations simultaneously, and the accelerated fluid might influence globally other reconnection sites. In studies with a single magnetic x-point, with open or periodic boundary conditions, the global dynamic is neglected. The approach here taken is more statistical in nature and the simulations are therefore structured as follows. 

First, in a $L_x\times L_y\times Lz = 1024\times1024\times1$ system  with periodic boundaries in $x$- and $z$-direction, a magnetic field of magnitude $10^{-4}$ (in numerical units) is imposed at the top ($y = 1024$) and bottom ($y = 1$) boundaries, in positive and negative $x-$direction, respectively. In this way, an external source of magnetic energy is introduced. Then, a time-dependent forcing induces turbulence in the plasma and drags the magnetic field throughout the system. Secondly, once the turbulence is fully developed, an algorithm identifies the diffusive regions, where opposed magnetic flux are pushed together by the turbulent motion. Finally, the out-of-plane electric field $E_z$ in the diffusive regions, as a measure for the reconnection rate $\Gamma = E_z/B_{\text{rms}}$, where $B_{\text{rms}} \equiv \sqrt{1/(L_xL_yL_z)\sum_{\vec{x}} \vec{B}^2}$ \citep{mohseni}, is analyzed and statistics is acquired. All the simulations have also been performed with different system sizes, and no difference was evident. 

All quantities will be denoted in numerical units. The simulations were performed with the following parameters: $\Delta t = \Delta x = \mu_0 = \epsilon_0 = 1$. The electronic and ionic charges and masses have been set to $q_0=-q_1 = -1$ and $m_0=1$, $m_1=100$, respectively, such that with the particle densities $n_0 = n_1=1$ they yield the mass densities $\rho_1=100\rho_0 = 100$. All the initial velocities are set to zero, while equal viscosities $\eta_0 = \eta_1 = 10^{-3}$ have been set via the parameters $\tau_0=\tau_1 = 0.503$.  For further details on the numerical model, see Ref. \citep{nostro_1}.

\subsection{Generating the turbulence}
\begin{figure}
		\centering
		\includegraphics[width=0.53\textwidth]{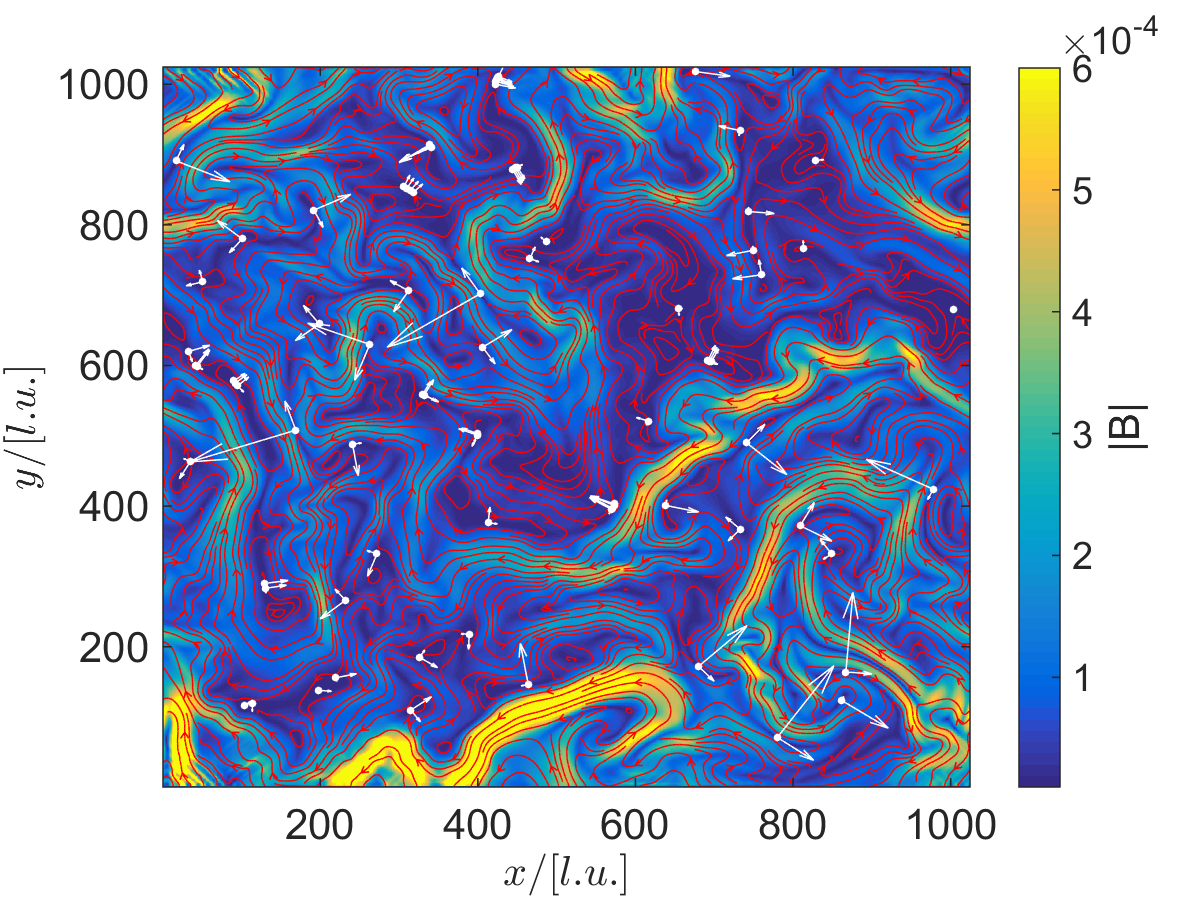}
		\caption{Snapshot of the system in turbulence. The colours denote the magnitude of $\vec{B}_S$ (see Sec. \ref{sec:rec_in_turb_plasma}), while  the red streamlines indicate the direction of the magnetic field. The white dots depict the reconnection sites identified by the algorithm described in Sec. \ref{sec:simulations}. For each reconnection site, the curvature along the depth and the length of the diffusive regions, as defined in Eq. \eqref{eq:depth_and_length}, are marked in white. For better visibility, the curvature along the length of the diffusive regions has been scaled by a factor of $5$.}
	\label{fig:magn_field}
\end{figure}
To ensure an isotropic turbulence, the force is given by a time-dependent superposition of force fields
\begin{equation} 
\label{eq:forces}
\vec{F}\kl{x} = \frac{A_0}{N}\sum_{l=1}^{N}{\vec{F}'_{l}\kl{\vec{x}}} \quad ,
\end{equation}
where $A_0=10^{-3}$ is the magnitude of each force, $N=10^3$ the number of superposed forces, and $\vec{F}'_l\kl{\vec{x}}$ is defined as
\begin{equation} 
\label{eq:sum_forces}
\begin{split}
F'_l\kl{x} &= \sin(k_{x,l} \bcdot x)\cos(k_{x,l}\bcdot y) \quad , \\
F'_l\kl{y} &= -\cos(k_{y,l}\bcdot x)\sin(k_{y,l}\bcdot y) \quad , \\
F'_l\kl{z} &= 0 \quad ,
\end{split}
\end{equation}
such that $\nabla \bcdot \vec{F}'\kl{l}  = 0$. The vectors $\vec{k}_l$ have constant length, chosen such that the force field has a wavelength of $32$ cells (or $32 \;\Delta x$). The wavelength of the force field is resolved by the numerical method. The direction of each of the $N$ vectors $\vec{k}_l$ is chosen from a uniform random distribution.  Every $10^4$ time steps, the distribution of $\vec{k}_l$ is randomly calculated again. According to \citet{boffetta}, the injection scale of the force can be seen in the power spectrum of the turbulence, where the slope of the inverse energy cascade changes to $-5/3$. This can be seen in Fig. \ref{fig:correlation}, and it occurs at $k \equiv |\vec{k}|\approx32$, which is the wavelength of the force field. The spectrum, as in Fig. \ref{fig:correlation}, remains steady throughout the measurements. The magnetic field of an exemplary time steps is depicted in Fig. \ref{fig:magn_field}. 
\begin{figure}
		\centering
		\includegraphics[width=0.5\textwidth]{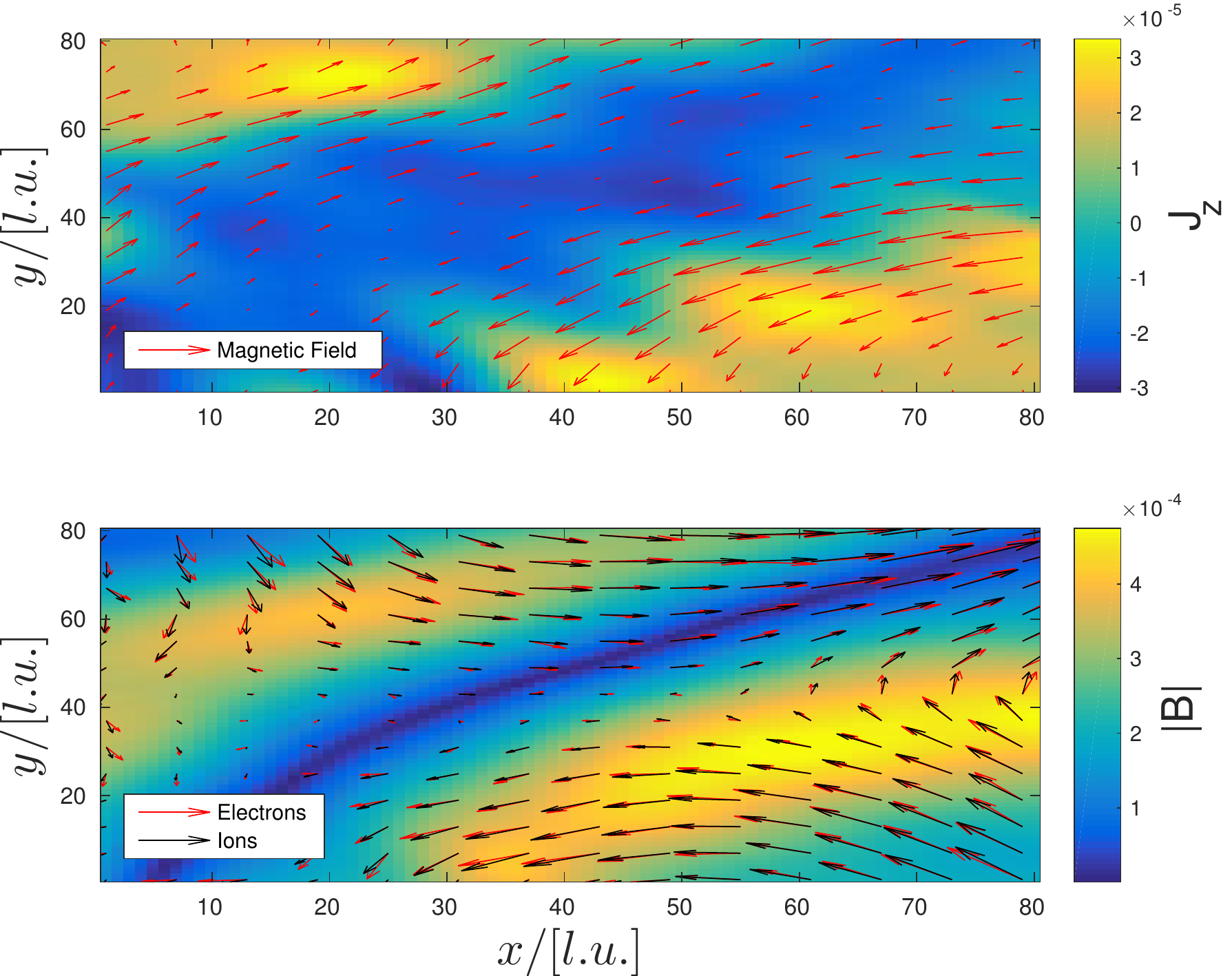}
		\caption{Close up of a reconnection site. On top, the out-of-plane current is color-coded, while in the lower panel the the color field denotes the magnitude of $\vec{B}_S$ (see Sec. \ref{sec:rec_in_turb_plasma}). In the lower panel, the velocity field of the electronic and ionic populations has been depicted, once the overall average velocity drift has been subtracted.}
	\label{fig:recon_site}
\end{figure}
To ensure that the collected data is indeed taken from uncorrelated samples, once the turbulence is developed, the behaviour of the fields were estimated by measuring the correlation function
\begin{equation}
\mathcal{C}_i\kl{t} \equiv \frac{1}{L_x L_y} \sum_{\vec{x}} s(B_i\ekl{\vec{x},t_0})\bcdot s(B_i\ekl{\vec{x},t_0+t}) \quad ,
\label{eq:corr_fun}
\end{equation}
where $s(B_i\ekl{\vec{x},t_0})$ is the sign of the $i$-component of the magnetic field at position $\vec{x}$ and time $t_0$. We assumed that the magnetic fields are uncorrelated once they fulfill  the condition
\begin{equation}
\mathcal{C}_i\kl{t} < \mathcal{SD}\kl{\mathcal{C}_i} \quad ,
\end{equation}
for $i=x,y$, where $SD\kl{\mathcal{C}_i}$ is the standard deviation of the natural fluctuations of the correlation function. As it can be seen in Fig. \ref{fig:correlation}, this happens about every $10^4$ time steps. Therefore, starting from $t=10^5 \Delta t$, when the turbulence is fully developed, measurements are taken every $2\times 10^4$ time steps until at least $t=3 \bcdot 10^5 \Delta t$. With Reynolds numbers $\text{Re}_s = L\bar{v}_s/\eta_s > 5\bcdot 10^{3}$, in all simulations we avoid vortex condensation.

\subsection{Reconnection rates in turbulent plasma}
\label{sec:rec_in_turb_plasma}
\begin{figure}
		\centering
		\includegraphics[width=0.5\textwidth]{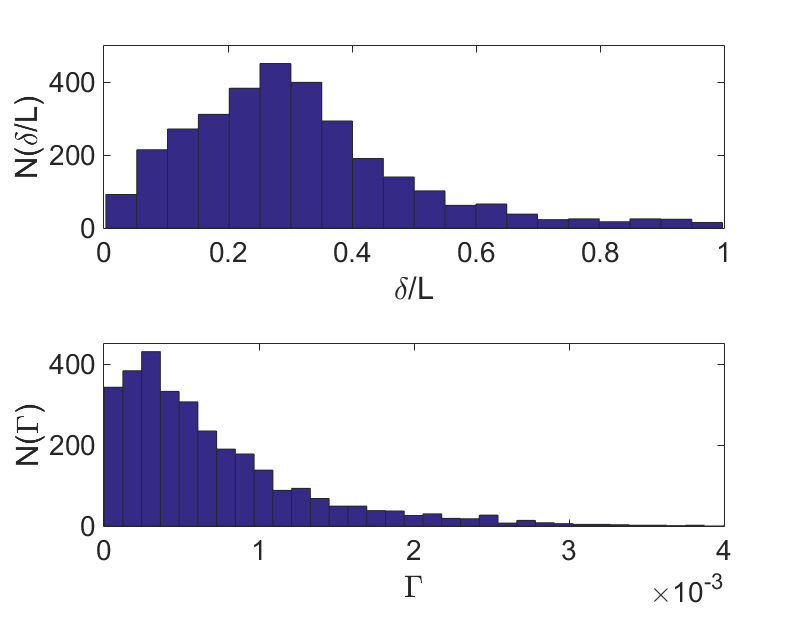}
		\caption{In the upper panel, the distribution of the aspect ratios $\delta/L = \sqrt{\lambda_2/\lambda_1}$ of the reconnection sites. In lower panel, the distribution of the measured reconnection rates. The peak of the of the reconnection events is around $5\bcdot 10^{-4}$ with a number of larger events that  seem to follow a power law distribution, although larger simulations are necessary to confirm this. }
	\label{fig:recon_site_stat}
\end{figure}
Magnetic reconnection occurs when two opposed magnetic flux tubes are brought together, as it can be seen in Fig. \ref{fig:recon_site}. Between them, i.e. inside the diffusive region, the magnetic field cancels out, and therefore, reconnection sites are characterized by a vanishing magnetic field. Considering the surface defined by $|\vec{B}_S|$, where $\vec{B}_S = (B_x, B_y) $, the nodes of the lattice where the condition
\begin{equation}
\label{eq:condition_cluster}
|\vec{B}_S| <  \,\frac{\epsilon}{L_x L_y} \sum _{\vec{x}}|\vec{B}_S| \quad ,
\end{equation}
is fulfilled are clustered with the Hoshen-Kopelman algorithm \citep{hoshen}, and the point of lowest $|\vec{B}_S|$ of each cluster is chosen. Note that the reconnection sites with a background guide field in $z$-direction are also taken into account. The results of the experiments are unaffected by the choice of $\epsilon$ in a range between $10^{-3}$ to $10^{-1}$. Data presented in this work have been gathered with $\epsilon = 1\bcdot 10^{-2}$.  To include only the magnetic x-points, and exclude eventual local maxima and minima of the surface defined by $|\vec{B}_S|$, we evaluate the curvature of the scalar potential defined by $ \nabla A \times \hat{e_z}\equiv\vec{B}_S $. Reconnection sites are saddle points of $A$ \citep{turb_2d},  and without the explicit knowledge of the magnetic potential, with its definition, the Hessian matrix
\begin{equation}
\label{eq:hessian}
\mathcal{H}^{i,j}_{ A }\kl{\vec{x}} = \pderDbl{A}{x_i}{x_j} \quad ,
\end{equation}
can be easily expressed with discrete derivatives of $B_x$ and $B_y$. If the eigenvalues are of opposite sign, the point is taken into account, discharged otherwise. If a cluster defined by Eq. \eqref{eq:condition_cluster} is divided by magnetic islands, as described in \citep{perspectives}, i.e. areas in which the eigenvalues of the Hessian are of the same sign, the cluster is divided into different reconnection sites accordingly.

We define $\lambda_1$ and $\lambda_2$ as the largest and smallest eigenvalue, respectively. The curvature is approximated with an ellipsoid, the axes of which are represented by the eigenvectors of the Hessian matrix. These are direct measures of the curvature at the sites, and therefore, the squared root of the inverse of the eigenvalues is proportional to the typical lengths along which the magnetic field changes. This means, the depth and the length of the diffusive regions can be expressed by
\begin{equation}
\label{eq:depth_and_length}
 \delta^{-2} \propto |\lambda_1| \quad ,  \quad
 L^{-2} \propto |\lambda_2| \quad .
\end{equation}
The eigenvalues differ roughly by a factor $10$, with variations that do not exceed an order of magnitude. The contained variations of the aspect ratio (see Fig. \ref{fig:recon_site_stat}) indicate that most of the reconnection sites have an aspect ratio around $\sim 0.3$, in contrast with other studies where this quantity can vary within several order of magnitudes \citep{yamada_two_fluid,turb_2d}.
\begin{figure}
		\centering
		\includegraphics[width=0.5\textwidth]{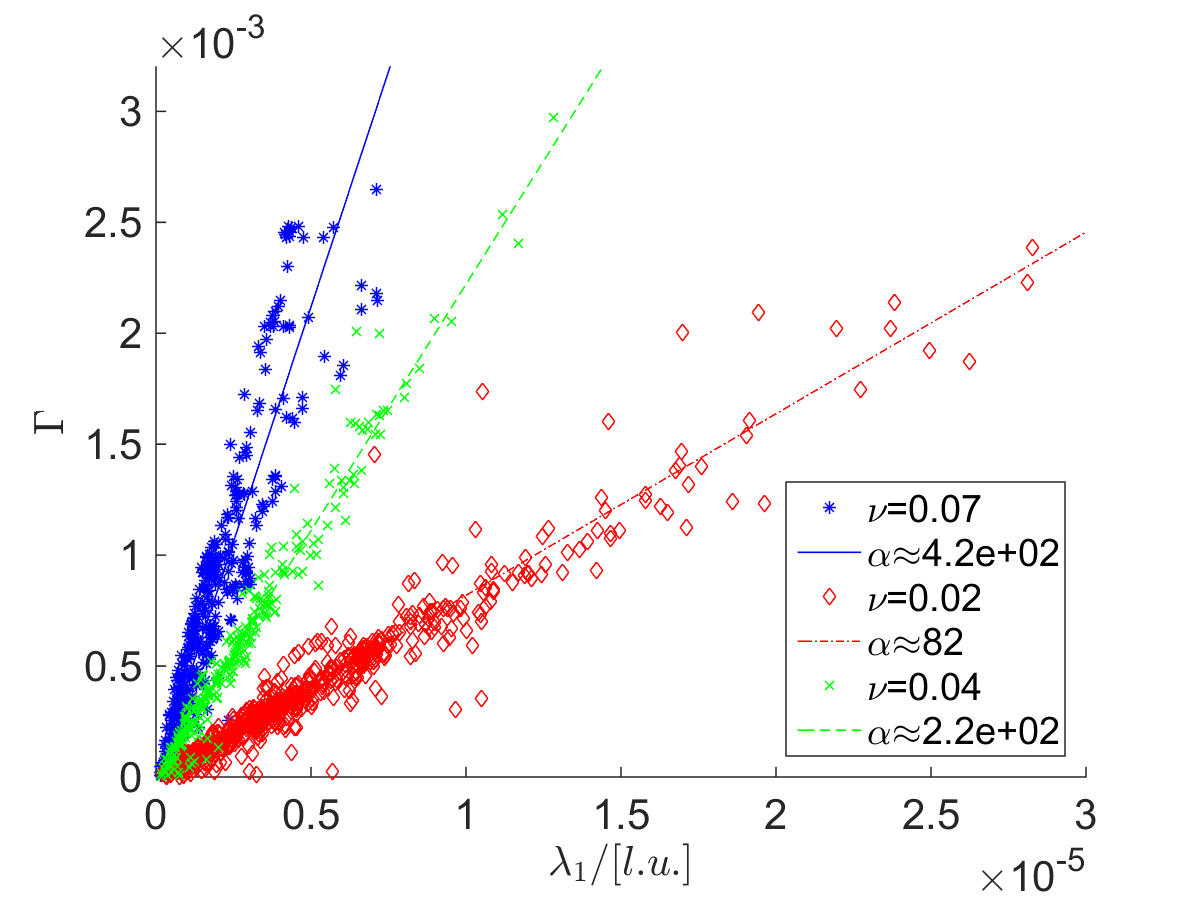}
		\caption{Relation between $\Gamma$, measured via the out-of-plane electric field $E_z$ \citep{mohseni}, and the largest eigenvalue $\lambda_1$ of the Hessian matrix, as defined in Eq. \eqref{eq:hessian}, for different collision frequencies $\nu$. Here, $\alpha$ denotes the slope of the fitting curves defined in Eq. \eqref{eq:slopes_1}.}
	\label{fig:lambda_corr}
\end{figure}
From Fig.~\ref{fig:recon_site_stat} we also observe that most of the reconnection events possess reconnection rates around $5\bcdot 10^{-4}$ with some larger values that seem to follow a power law distribution, although larger simulations will be needed to confirm the latter statement. 

Another interesting finding is that the reconnection rates $\Gamma = E_z/B_{\text{rms}}$ depend linearly on the largest eigenvalue of the Hessian matrix. See Fig. \ref{fig:lambda_corr} where three examples for three different collision frequencies are shown, fitted with polynomial of first degree:
\begin{equation}
 \label{eq:slopes_1}
\Gamma = \alpha\kl{\nu} \lambda_1 \quad ,
\end{equation}
with $\alpha\kl{\nu}$ a fitting parameter. We have also noticed that $\delta$ and $L$ are loosely correlated, and therefore, the reconnection rates might be written as proportional to the inverse of the area of the diffusive region, and not to their aspect ratio. In fact, we verified that $\delta/L$ and $\Gamma$ show no correlation throughout our simulations.  

\begin{figure}
		\centering
		\includegraphics[width=0.5\textwidth]{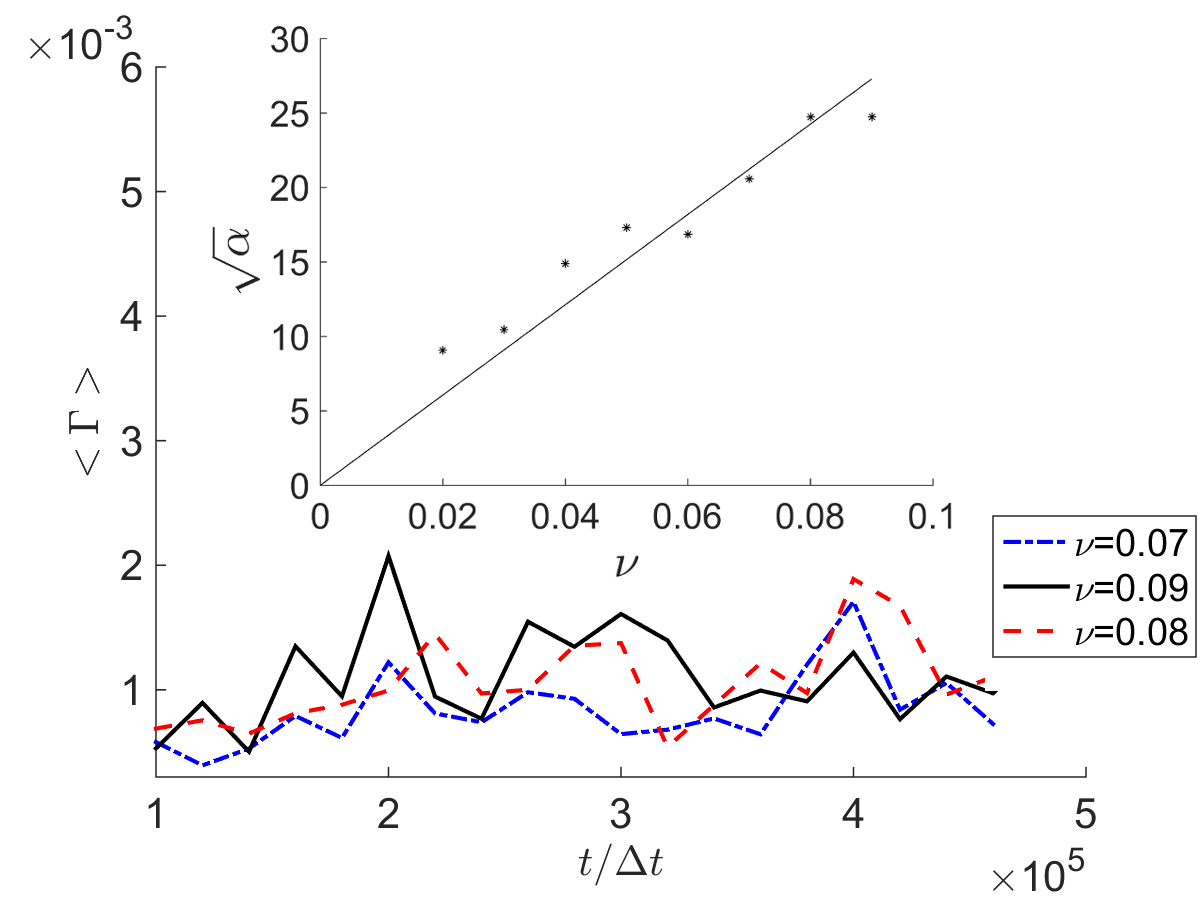}
		\caption{ Example of the time evolution of the average reconnection rate $\Gamma$ for three different $\nu$. During the simulations the rates fluctuate around a constant average. In the upper panel, the correlation coefficient $\alpha$ (slopes in Fig. \ref{fig:lambda_corr}) versus the collision frequency $\nu$.  }
	\label{fig:slope_corr}
\end{figure}

By tuning $\nu$, we found that the slope of the correlation between the reconnection rates and $\lambda_1$ scales with the collision frequency squared (See Fig. \ref{fig:slope_corr}). That is, the parameter $\alpha\kl{\nu}$ in Eq. \eqref{eq:slopes_1} can also be expressed as:
\begin{equation}
 \label{eq:slopes_2}
\sqrt{\alpha} = \beta \nu \quad ,
\end{equation}
with $\beta$ a fitting parameter. Finally, the reconnection rate using Eqs.~\eqref{eq:depth_and_length}, \eqref{eq:slopes_1}, and \eqref{eq:slopes_2} can be rewritten as 
\begin{equation}
\label{eq:correlations_here}
\Gamma \propto \nu^2 \lambda_1 \propto \frac{1}{\sigma^2\delta^2} \quad .
\end{equation}

As mentioned in Sec. \ref{sp_model}, the reconnection rate in the Sweet-Parker model is thought to be inversely proportional to the conductivity, and to the inverse of its logarithm in the Petschek model. For a comparison of the Sweet-Parker, $\Gamma_{\text{SP}}$ in Eq. \eqref{eq:gamma_from_fields}, and Petschek reconnection rates, $\Gamma_{\text{P}}$ in Eq. \eqref{eq:petsch_rate}, to those measured in our simulations, $\Gamma$, the dependencies can be summarized as follows:
\begin{equation} 
\label{eq:sw_and_p}
\begin{split}
\Gamma_{\text{SP}} &\propto {\frac{1}{\sigma\delta}}\propto\nu\sqrt{\lambda_1} \quad , \\
\Gamma_{\text{P}} &\propto \frac{1}{\log{\kl{ \sigma\delta}}}\propto \frac{-1}{\log{\kl{\nu\sqrt{\lambda_1}}}} \quad , \\
\Gamma &\propto \frac{1}{\sigma^2\delta^2}\propto  \nu^2 \lambda_1 \quad .\\
\end{split}
\end{equation}

A final observation is that the ratio $\delta/L = \sqrt{{\lambda_2}/{\lambda_1}}$ is not affected by changing the collision frequency. This indicates that the geometry of the reconnection sites is  independent from the parameter $\nu$, although the collision frequency affects the reconnection rates and the size of the diffusive regions. Note also that the average reconnection rate remains almost constant in time throughout the whole simulation (see Fig.~\ref{fig:slope_corr}).

\section{\label{sec:concl}Conclusions and Outlook}
We have performed simulations of a two-fluid turbulent plasma and studied the magnetic reconnection process. By varying the electron-ion collision frequency, and therefore the conductivity of the plasma, we have measured a new relation between the reconnection rate, the plasma conductivity and the depth of the diffusive region, namely, 
\begin{equation}
\label{eq:final}
\Gamma \propto \frac{1}{\sigma^2\delta^2} \quad .
\end{equation}
The reconnection rates measured lied within the orders of $10^{-7}-10^{-3}$. Experimental studies with variational electron-ion collisionality  \citep{yamada_two_fluid} seem to corroborate the fact that the reconnection rate is affected by the collision frequency. To our knowledge, the reconnection rate in a turbulent plasma has not yet been measured experimentally, and therefore Eq. \eqref{eq:final} could not be validated. 

By comparing our simulations to the Sweet-Parker and Petschek models, we evince some discrepancies, in particular in the proportionality of the reconnection rates to the plasma conductivity, $\propto 1/\sigma$ (Sweet-Parker) and $\propto1/\log{\sigma}$ (Petschek). Also the measured aspect ratios with values around $0.3$ rule out the characteristic Sweet-Parker geometry which assumes $\delta \ll L$. The reason for this discrepancies can be due to the fact that we considered a turbulent two-species plasma that satisfies a generalized Ohm's law, while the mentioned theoretical models are based on time-independent resistive magnetohydrodynamics, where the origin of the diffusive region is not well-defined. In future, it would be interesting to expand the simulations  of turbulent two-fluid plasmas from the 2.5D used in this work to a fully three-dimensional system.

Our results suggest that the aspect ratio of the diffusive region is unaffected by changing the collision frequency within one order of magnitude. In over 98\% of the cases, $\delta/L$ assumed values between $0.1$ and $0.9$, with a peak around $0.3$, and no correlation between $\delta/L$ and $\nu$ could made be evident. 
Another interesting finding, is that $\delta/L$ does not seem to affect the reconnection rates, in contrast with other results based on MHD \cite{turb_2d}. Since $\delta$ and $L$ are correlated, our reconnection rate can also be expressed as proportional to the inverse of the area of the diffusive region, rather than to its aspect ratio. 

Changes in the global Reynolds number, although limited to one order of magnitude due to stability reasons, caused no noticeable effects on our findings. The dependence of the aspect ratio with respect to the Reynolds number of the system, as well as other plasma characteristic quantities, is a question that we intend to address in future. Additionally, as data from the MMS  \citep{burch_magnetospheric} will be made available, we expect to compare our results to real physical systems.

\begin{acknowledgments}
Financial support from the European Research Council (ERC) Advanced Grant 319968-FlowCCS is kindly acknowledged.
\end{acknowledgments}

\bibliography{literature}

\begin{thebibliography}{29}%
\makeatletter
\providecommand \@ifxundefined [1]{%
 \@ifx{#1\undefined}
}%
\providecommand \@ifnum [1]{%
 \ifnum #1\expandafter \@firstoftwo
 \else \expandafter \@secondoftwo
 \fi
}%
\providecommand \@ifx [1]{%
 \ifx #1\expandafter \@firstoftwo
 \else \expandafter \@secondoftwo
 \fi
}%
\providecommand \natexlab [1]{#1}%
\providecommand \enquote  [1]{``#1''}%
\providecommand \bibnamefont  [1]{#1}%
\providecommand \bibfnamefont [1]{#1}%
\providecommand \citenamefont [1]{#1}%
\providecommand \href@noop [0]{\@secondoftwo}%
\providecommand \href [0]{\begingroup \@sanitize@url \@href}%
\providecommand \@href[1]{\@@startlink{#1}\@@href}%
\providecommand \@@href[1]{\endgroup#1\@@endlink}%
\providecommand \@sanitize@url [0]{\catcode `\\12\catcode `\$12\catcode
  `\&12\catcode `\#12\catcode `\^12\catcode `\_12\catcode `\%12\relax}%
\providecommand \@@startlink[1]{}%
\providecommand \@@endlink[0]{}%
\providecommand \url  [0]{\begingroup\@sanitize@url \@url }%
\providecommand \@url [1]{\endgroup\@href {#1}{\urlprefix }}%
\providecommand \urlprefix  [0]{URL }%
\providecommand \Eprint [0]{\href }%
\providecommand \doibase [0]{http://dx.doi.org/}%
\providecommand \selectlanguage [0]{\@gobble}%
\providecommand \bibinfo  [0]{\@secondoftwo}%
\providecommand \bibfield  [0]{\@secondoftwo}%
\providecommand \translation [1]{[#1]}%
\providecommand \BibitemOpen [0]{}%
\providecommand \bibitemStop [0]{}%
\providecommand \bibitemNoStop [0]{.\EOS\space}%
\providecommand \EOS [0]{\spacefactor3000\relax}%
\providecommand \BibitemShut  [1]{\csname bibitem#1\endcsname}%
\let\auto@bib@innerbib\@empty
\bibitem [{\citenamefont {Munsat}\ \emph {et~al.}(2007)\citenamefont {Munsat},
  \citenamefont {Park}, \citenamefont {Classen}, \citenamefont {Domier},
  \citenamefont {Donne}, \citenamefont {Luhmann}, \citenamefont {Mazzucato},
  \citenamefont {Van~de Pol},\ and\ \citenamefont {team}}]{MR_in_tokamak}%
  \BibitemOpen
  \bibfield  {author} {\bibinfo {author} {\bibfnamefont {M.}~\bibnamefont
  {Munsat}}, \bibinfo {author} {\bibfnamefont {H.}~\bibnamefont {Park}},
  \bibinfo {author} {\bibfnamefont {I.}~\bibnamefont {Classen}}, \bibinfo
  {author} {\bibfnamefont {C.}~\bibnamefont {Domier}}, \bibinfo {author}
  {\bibfnamefont {A.}~\bibnamefont {Donne}}, \bibinfo {author} {\bibfnamefont
  {N.}~\bibnamefont {Luhmann}, \bibfnamefont {Jr}}, \bibinfo {author}
  {\bibfnamefont {E.}~\bibnamefont {Mazzucato}}, \bibinfo {author}
  {\bibfnamefont {M.}~\bibnamefont {Van~de Pol}}, \ and\ \bibinfo {author}
  {\bibfnamefont {T.}~\bibnamefont {team}},\ }\href
  {http://stacks.iop.org/0029-5515/47/i=11/a=L01} {\bibfield  {journal}
  {\bibinfo  {journal} {Nuclear Fusion}\ } (\bibinfo {year}
  {2007})}\BibitemShut {NoStop}%
\bibitem [{\citenamefont {Gonzalez}\ and\ \citenamefont
  {Parker}(2016)}]{gonzalez2016magnetic}%
  \BibitemOpen
  \bibfield  {author} {\bibinfo {author} {\bibfnamefont {W.}~\bibnamefont
  {Gonzalez}}\ and\ \bibinfo {author} {\bibfnamefont {E.}~\bibnamefont
  {Parker}},\ }\href {https://books.google.ch/books?id=R2OhCwAAQBAJ} {\emph
  {\bibinfo {title} {Magnetic Reconnection: Concepts and Applications}}},\
  Astrophysics and Space Science Library\ (\bibinfo  {publisher} {Springer
  International Publishing},\ \bibinfo {year} {2016})\BibitemShut {NoStop}%
\bibitem [{\citenamefont {Priest}\ and\ \citenamefont {Forbes}(2000)}]{priest}%
  \BibitemOpen
  \bibfield  {author} {\bibinfo {author} {\bibfnamefont {E.}~\bibnamefont
  {Priest}}\ and\ \bibinfo {author} {\bibfnamefont {T.}~\bibnamefont
  {Forbes}},\ }\href@noop {} {\bibfield  {journal} {\bibinfo  {journal}
  {Cambridge University Press}\ } (\bibinfo {year} {2000})}\BibitemShut
  {NoStop}%
\bibitem [{\citenamefont {Veronig}\ and\ \citenamefont
  {Polanec}(2015)}]{energy}%
  \BibitemOpen
  \bibfield  {author} {\bibinfo {author} {\bibfnamefont {A.~M.}\ \bibnamefont
  {Veronig}}\ and\ \bibinfo {author} {\bibfnamefont {W.}~\bibnamefont
  {Polanec}},\ }\href {\doibase 10.1007/s11207-015-0789-6} {\bibfield
  {journal} {\bibinfo  {journal} {Solar Physics}\ }\textbf {\bibinfo {volume}
  {290}},\ \bibinfo {pages} {2923} (\bibinfo {year} {2015})}\BibitemShut
  {NoStop}%
\bibitem [{\citenamefont {Zweibel}\ and\ \citenamefont
  {Yamada}(2009)}]{yamada_rev_1}%
  \BibitemOpen
  \bibfield  {author} {\bibinfo {author} {\bibfnamefont {E.~G.}\ \bibnamefont
  {Zweibel}}\ and\ \bibinfo {author} {\bibfnamefont {M.}~\bibnamefont
  {Yamada}},\ }\href {\doibase 10.1146/annurev-astro-082708-101726} {\bibfield
  {journal} {\bibinfo  {journal} {Annual Review of Astronomy and Astrophysics}\
  }\textbf {\bibinfo {volume} {47}},\ \bibinfo {pages} {291} (\bibinfo {year}
  {2009})},\ \Eprint
  {http://arxiv.org/abs/http://dx.doi.org/10.1146/annurev-astro-082708-101726}
  {http://dx.doi.org/10.1146/annurev-astro-082708-101726} \BibitemShut
  {NoStop}%
\bibitem [{\citenamefont {Yamada}\ \emph {et~al.}(2010)\citenamefont {Yamada},
  \citenamefont {Masaaki}, \citenamefont {Kulsrud}, \citenamefont {Russell},\
  and\ \citenamefont {Ji}}]{yamada_rev_2}%
  \BibitemOpen
  \bibfield  {author} {\bibinfo {author} {\bibnamefont {Yamada}}, \bibinfo
  {author} {\bibnamefont {Masaaki}}, \bibinfo {author} {\bibnamefont
  {Kulsrud}}, \bibinfo {author} {\bibnamefont {Russell}}, \ and\ \bibinfo
  {author} {\bibfnamefont {H.}~\bibnamefont {Ji}},\ }\href {\doibase
  10.1103/RevModPhys.82.603} {\bibfield  {journal} {\bibinfo  {journal} {Rev.
  Mod. Phys.}\ }\textbf {\bibinfo {volume} {82}},\ \bibinfo {pages} {603}
  (\bibinfo {year} {2010})}\BibitemShut {NoStop}%
\bibitem [{\citenamefont {Yamada}\ \emph {et~al.}(2006)\citenamefont {Yamada},
  \citenamefont {Ren}, \citenamefont {Ji}, \citenamefont {Breslau},
  \citenamefont {Gerhardt}, \citenamefont {Kulsrud},\ and\ \citenamefont
  {Kuritsyn}}]{yamada_two_fluid}%
  \BibitemOpen
  \bibfield  {author} {\bibinfo {author} {\bibfnamefont {M.}~\bibnamefont
  {Yamada}}, \bibinfo {author} {\bibfnamefont {Y.}~\bibnamefont {Ren}},
  \bibinfo {author} {\bibfnamefont {H.}~\bibnamefont {Ji}}, \bibinfo {author}
  {\bibfnamefont {J.}~\bibnamefont {Breslau}}, \bibinfo {author} {\bibfnamefont
  {S.}~\bibnamefont {Gerhardt}}, \bibinfo {author} {\bibfnamefont
  {R.}~\bibnamefont {Kulsrud}}, \ and\ \bibinfo {author} {\bibfnamefont
  {A.}~\bibnamefont {Kuritsyn}},\ }\href {\doibase
  http://dx.doi.org/10.1063/1.2203950} {\bibfield  {journal} {\bibinfo
  {journal} {Physics of Plasmas}\ }\textbf {\bibinfo {volume} {13}},\ \bibinfo
  {eid} {052119} (\bibinfo {year} {2006}),\
  http://dx.doi.org/10.1063/1.2203950}\BibitemShut {NoStop}%
\bibitem [{\citenamefont {Lewis}\ \emph {et~al.}(2014)\citenamefont {Lewis},
  \citenamefont {Antiochos},\ and\ \citenamefont {Drake}}]{perspectives}%
  \BibitemOpen
  \bibfield  {author} {\bibinfo {author} {\bibfnamefont {W.}~\bibnamefont
  {Lewis}}, \bibinfo {author} {\bibfnamefont {S.}~\bibnamefont {Antiochos}}, \
  and\ \bibinfo {author} {\bibfnamefont {J.}~\bibnamefont {Drake}},\
  }\href@noop {} {\emph {\bibinfo {title} {Magnetic Reconnection: Theoretical
  and Observational Perspectives}}}\ (\bibinfo  {publisher} {Springer New
  York},\ \bibinfo {year} {2014})\BibitemShut {NoStop}%
\bibitem [{\citenamefont {Daughton}\ \emph {et~al.}(2011)\citenamefont
  {Daughton}, \citenamefont {Roytershteyn}, \citenamefont {Karimabadi},
  \citenamefont {Yin}, \citenamefont {Albright}, \citenamefont {Bergen},\ and\
  \citenamefont {Bowers}}]{3D}%
  \BibitemOpen
  \bibfield  {author} {\bibinfo {author} {\bibfnamefont {W.}~\bibnamefont
  {Daughton}}, \bibinfo {author} {\bibfnamefont {V.}~\bibnamefont
  {Roytershteyn}}, \bibinfo {author} {\bibfnamefont {H.}~\bibnamefont
  {Karimabadi}}, \bibinfo {author} {\bibfnamefont {L.}~\bibnamefont {Yin}},
  \bibinfo {author} {\bibfnamefont {B.}~\bibnamefont {Albright}}, \bibinfo
  {author} {\bibfnamefont {B.}~\bibnamefont {Bergen}}, \ and\ \bibinfo {author}
  {\bibfnamefont {K.~J.}\ \bibnamefont {Bowers}},\ }\href
  {http://www.nature.com/nphys/journal/v7/n7/full/nphys1965.html} {\bibfield
  {journal} {\bibinfo  {journal} {Nat. Phys.}\ } (\bibinfo {year}
  {2011})}\BibitemShut {NoStop}%
\bibitem [{\citenamefont {Sweet}(1958)}]{sweet}%
  \BibitemOpen
  \bibfield  {author} {\bibinfo {author} {\bibfnamefont {P.}~\bibnamefont
  {Sweet}},\ }in\ \href@noop {} {\emph {\bibinfo {booktitle} {Electromagnetic
  Phenomena in Cosmical Physics}}},\ \bibinfo {series} {IAU Symposium},
  Vol.~\bibinfo {volume} {6},\ \bibinfo {editor} {edited by\ \bibinfo {editor}
  {\bibfnamefont {B.}~\bibnamefont {{Lehnert}}}}\ (\bibinfo {year} {1958})\ p.\
  \bibinfo {pages} {123}\BibitemShut {NoStop}%
\bibitem [{\citenamefont {Parker}(1957)}]{parker}%
  \BibitemOpen
  \bibfield  {author} {\bibinfo {author} {\bibfnamefont {E.~N.}\ \bibnamefont
  {Parker}},\ }\href {\doibase 10.1029/JZ062i004p00509} {\bibfield  {journal}
  {\bibinfo  {journal} {Journal of Geophysical Research}\ }\textbf {\bibinfo
  {volume} {62}},\ \bibinfo {pages} {509} (\bibinfo {year} {1957})}\BibitemShut
  {NoStop}%
\bibitem [{\citenamefont {Petscheck}(1964)}]{petscheck}%
  \BibitemOpen
  \bibfield  {author} {\bibinfo {author} {\bibfnamefont {H.}~\bibnamefont
  {Petscheck}},\ }\href@noop {} {\bibfield  {journal} {\bibinfo  {journal}
  {Proceedings of the AAS-NASA Symposium}\ } (\bibinfo {year}
  {1964})}\BibitemShut {NoStop}%
\bibitem [{\citenamefont {Retino}\ \emph {et~al.}(2006)\citenamefont {Retino},
  \citenamefont {Sundkvist}, \citenamefont {Vaivads}, \citenamefont {Mozer},\
  and\ \citenamefont {Owen}}]{modern_turbulence}%
  \BibitemOpen
  \bibfield  {author} {\bibinfo {author} {\bibfnamefont {A.}~\bibnamefont
  {Retino}}, \bibinfo {author} {\bibfnamefont {D.}~\bibnamefont {Sundkvist}},
  \bibinfo {author} {\bibfnamefont {A.}~\bibnamefont {Vaivads}}, \bibinfo
  {author} {\bibfnamefont {F.}~\bibnamefont {Mozer}}, \ and\ \bibinfo {author}
  {\bibfnamefont {J.}~\bibnamefont {Owen}},\ }\href {\doibase 10.1038/nphys574}
  {\bibfield  {journal} {\bibinfo  {journal} {Nat. Phys.}\ }\textbf {\bibinfo
  {volume} {3}},\ \bibinfo {pages} {235} (\bibinfo {year} {2006})}\BibitemShut
  {NoStop}%
\bibitem [{\citenamefont {Dorelli}\ and\ \citenamefont
  {Bhattacharjee}(2008)}]{sim_earth_magnetosph}%
  \BibitemOpen
  \bibfield  {author} {\bibinfo {author} {\bibfnamefont {J.~C.}\ \bibnamefont
  {Dorelli}}\ and\ \bibinfo {author} {\bibfnamefont {A.}~\bibnamefont
  {Bhattacharjee}},\ }\href {\doibase 10.1063/1.2913548} {\bibfield  {journal}
  {\bibinfo  {journal} {Physics of Plasmas}\ }\textbf {\bibinfo {volume}
  {15}},\ \bibinfo {pages} {056504} (\bibinfo {year} {2008})},\ \Eprint
  {http://arxiv.org/abs/http://dx.doi.org/10.1063/1.2913548}
  {http://dx.doi.org/10.1063/1.2913548} \BibitemShut {NoStop}%
\bibitem [{\citenamefont {Burch}\ and\ \citenamefont
  {Torbert}(2016)}]{burch_magnetospheric}%
  \BibitemOpen
  \bibfield  {author} {\bibinfo {author} {\bibfnamefont {J.}~\bibnamefont
  {Burch}}\ and\ \bibinfo {author} {\bibfnamefont {R.}~\bibnamefont
  {Torbert}},\ }\href {https://books.google.de/books?id=pr69jwEACAAJ} {\emph
  {\bibinfo {title} {Magnetospheric Multiscale: A Mission to Investigate the
  Physics of Magnetic Reconnection}}}\ (\bibinfo  {publisher} {Springer
  Netherlands},\ \bibinfo {year} {2016})\BibitemShut {NoStop}%
\bibitem [{\citenamefont {Loureiro}\ \emph {et~al.}(2009)\citenamefont
  {Loureiro}, \citenamefont {Uzdensky}, \citenamefont {Schekochihin},
  \citenamefont {Cowley},\ and\ \citenamefont {Yousef}}]{Loureiro}%
  \BibitemOpen
  \bibfield  {author} {\bibinfo {author} {\bibfnamefont {N.~F.}\ \bibnamefont
  {Loureiro}}, \bibinfo {author} {\bibfnamefont {D.~A.}\ \bibnamefont
  {Uzdensky}}, \bibinfo {author} {\bibfnamefont {A.~A.}\ \bibnamefont
  {Schekochihin}}, \bibinfo {author} {\bibfnamefont {S.~C.}\ \bibnamefont
  {Cowley}}, \ and\ \bibinfo {author} {\bibfnamefont {T.~A.}\ \bibnamefont
  {Yousef}},\ }\href {\doibase 10.1111/j.1745-3933.2009.00742.x} {\bibfield
  {journal} {\bibinfo  {journal} {Monthly Notices of the Royal Astronomical
  Society: Letters}\ }\textbf {\bibinfo {volume} {399}},\ \bibinfo {pages}
  {L146} (\bibinfo {year} {2009})}\BibitemShut {NoStop}%
\bibitem [{\citenamefont {Servidio}\ \emph {et~al.}(2009)\citenamefont
  {Servidio}, \citenamefont {Matthaeus}, \citenamefont {Shay}, \citenamefont
  {Cassak},\ and\ \citenamefont {Dmitruk}}]{turb_2d}%
  \BibitemOpen
  \bibfield  {author} {\bibinfo {author} {\bibfnamefont {S.}~\bibnamefont
  {Servidio}}, \bibinfo {author} {\bibfnamefont {W.~H.}\ \bibnamefont
  {Matthaeus}}, \bibinfo {author} {\bibfnamefont {M.~A.}\ \bibnamefont {Shay}},
  \bibinfo {author} {\bibfnamefont {P.~A.}\ \bibnamefont {Cassak}}, \ and\
  \bibinfo {author} {\bibfnamefont {P.}~\bibnamefont {Dmitruk}},\ }\href
  {\doibase 10.1103/PhysRevLett.102.115003} {\bibfield  {journal} {\bibinfo
  {journal} {Phys. Rev. Lett.}\ }\textbf {\bibinfo {volume} {102}},\ \bibinfo
  {pages} {115003} (\bibinfo {year} {2009})}\BibitemShut {NoStop}%
\bibitem [{\citenamefont {Offeddu}\ and\ \citenamefont
  {Mendoza}(2016)}]{nostro_1}%
  \BibitemOpen
  \bibfield  {author} {\bibinfo {author} {\bibfnamefont {N.}~\bibnamefont
  {Offeddu}}\ and\ \bibinfo {author} {\bibfnamefont {M.}~\bibnamefont
  {Mendoza}},\ }\href {https://arxiv.org/abs/1612.06208} {\bibfield  {journal}
  {\bibinfo  {journal} {preprint Arxiv}\ } (\bibinfo {year} {2016})},\ \Eprint
  {http://arxiv.org/abs/https://arxiv.org/abs/1612.06208}
  {https://arxiv.org/abs/1612.06208} \BibitemShut {NoStop}%
\bibitem [{\citenamefont {Chen}\ \emph {et~al.}(1991)\citenamefont {Chen},
  \citenamefont {Chen}, \citenamefont {Martnez},\ and\ \citenamefont
  {Matthaeus}}]{lb_mhd_vecchio}%
  \BibitemOpen
  \bibfield  {author} {\bibinfo {author} {\bibfnamefont {S.}~\bibnamefont
  {Chen}}, \bibinfo {author} {\bibfnamefont {H.}~\bibnamefont {Chen}}, \bibinfo
  {author} {\bibfnamefont {D.}~\bibnamefont {Martnez}}, \ and\ \bibinfo
  {author} {\bibfnamefont {W.}~\bibnamefont {Matthaeus}},\ }\href {\doibase
  10.1103/PhysRevLett.67.3776} {\bibfield  {journal} {\bibinfo  {journal}
  {Phys. Rev. Lett.}\ }\textbf {\bibinfo {volume} {67}},\ \bibinfo {pages}
  {3776} (\bibinfo {year} {1991})}\BibitemShut {NoStop}%
\bibitem [{\citenamefont {Mendoza}\ and\ \citenamefont
  {Mu\~noz}(2010)}]{miller_2}%
  \BibitemOpen
  \bibfield  {author} {\bibinfo {author} {\bibfnamefont {M.}~\bibnamefont
  {Mendoza}}\ and\ \bibinfo {author} {\bibfnamefont {J.~D.}\ \bibnamefont
  {Mu\~noz}},\ }\href {\doibase 10.1103/PhysRevE.82.056708} {\bibfield
  {journal} {\bibinfo  {journal} {Phys. Rev. E}\ }\textbf {\bibinfo {volume}
  {82}},\ \bibinfo {pages} {056708} (\bibinfo {year} {2010})}\BibitemShut
  {NoStop}%
\bibitem [{\citenamefont {Alfv\'en}(1943)}]{frozen_in}%
  \BibitemOpen
  \bibfield  {author} {\bibinfo {author} {\bibfnamefont {H.}~\bibnamefont
  {Alfv\'en}},\ }\href@noop {} {\bibfield  {journal} {\bibinfo  {journal}
  {Arkiv f. Mat., Astron. o. Fys.}\ } (\bibinfo {year} {1943})}\BibitemShut
  {NoStop}%
\bibitem [{\citenamefont {Matthaeus}\ and\ \citenamefont
  {Velli}(2011)}]{turbulence_1}%
  \BibitemOpen
  \bibfield  {author} {\bibinfo {author} {\bibfnamefont {W.}~\bibnamefont
  {Matthaeus}}\ and\ \bibinfo {author} {\bibfnamefont {M.}~\bibnamefont
  {Velli}},\ }\href@noop {} {\bibfield  {journal} {\bibinfo  {journal} {Space
  Sci. Rev}\ } (\bibinfo {year} {2011})}\BibitemShut {NoStop}%
\bibitem [{\citenamefont {Uzdensky}\ and\ \citenamefont
  {Rightley}(2014)}]{uzdensky}%
  \BibitemOpen
  \bibfield  {author} {\bibinfo {author} {\bibfnamefont {D.}~\bibnamefont
  {Uzdensky}}\ and\ \bibinfo {author} {\bibfnamefont {S.}~\bibnamefont
  {Rightley}},\ }\href@noop {} {\bibfield  {journal} {\bibinfo  {journal}
  {Arxiv.org}\ } (\bibinfo {year} {2014})}\BibitemShut {NoStop}%
\bibitem [{Som(2007)}]{Somov2007}%
  \BibitemOpen
  \enquote {\bibinfo {title} {The generalized ohm's law in plasma},}\ in\ \href
  {\doibase 10.1007/978-0-387-68894-7_12} {\emph {\bibinfo {booktitle} {Plasma
  Astrophysics}}}\ (\bibinfo  {publisher} {Springer New York},\ \bibinfo
  {address} {New York, NY},\ \bibinfo {year} {2007})\ pp.\ \bibinfo {pages}
  {193--204}\BibitemShut {NoStop}%
\bibitem [{\citenamefont {Fridman}\ and\ \citenamefont
  {Kennedy}(2004)}]{fridman}%
  \BibitemOpen
  \bibfield  {author} {\bibinfo {author} {\bibfnamefont {A.}~\bibnamefont
  {Fridman}}\ and\ \bibinfo {author} {\bibfnamefont {L.}~\bibnamefont
  {Kennedy}},\ }\href@noop {} {\emph {\bibinfo {title} {Plasma Physics and
  Engineering}}}\ (\bibinfo  {publisher} {Taylor \& Francis},\ \bibinfo {year}
  {2004})\BibitemShut {NoStop}%
\bibitem [{\citenamefont {Mendoza}\ and\ \citenamefont
  {Mu\~noz}(2008)}]{miller_1}%
  \BibitemOpen
  \bibfield  {author} {\bibinfo {author} {\bibfnamefont {M.}~\bibnamefont
  {Mendoza}}\ and\ \bibinfo {author} {\bibfnamefont {J.~D.}\ \bibnamefont
  {Mu\~noz}},\ }\href {\doibase 10.1103/PhysRevE.77.026713} {\bibfield
  {journal} {\bibinfo  {journal} {Phys. Rev. E}\ }\textbf {\bibinfo {volume}
  {77}},\ \bibinfo {pages} {026713} (\bibinfo {year} {2008})}\BibitemShut
  {NoStop}%
\bibitem [{\citenamefont {Boffetta}\ and\ \citenamefont
  {Ecke}(2012)}]{boffetta}%
  \BibitemOpen
  \bibfield  {author} {\bibinfo {author} {\bibfnamefont {G.}~\bibnamefont
  {Boffetta}}\ and\ \bibinfo {author} {\bibfnamefont {R.~E.}\ \bibnamefont
  {Ecke}},\ }\href {\doibase 10.1146/annurev-fluid-120710-101240} {\bibfield
  {journal} {\bibinfo  {journal} {Annual Review of Fluid Mechanics}\ }\textbf
  {\bibinfo {volume} {44}},\ \bibinfo {pages} {427} (\bibinfo {year} {2012})},\
  \Eprint
  {http://arxiv.org/abs/http://dx.doi.org/10.1146/annurev-fluid-120710-101240}
  {http://dx.doi.org/10.1146/annurev-fluid-120710-101240} \BibitemShut
  {NoStop}%
\bibitem [{\citenamefont {Mohseni}\ \emph {et~al.}(2015)\citenamefont
  {Mohseni}, \citenamefont {Mendoza}, \citenamefont {Succi},\ and\
  \citenamefont {Herrmann}}]{mohseni}%
  \BibitemOpen
  \bibfield  {author} {\bibinfo {author} {\bibfnamefont {F.}~\bibnamefont
  {Mohseni}}, \bibinfo {author} {\bibfnamefont {M.}~\bibnamefont {Mendoza}},
  \bibinfo {author} {\bibfnamefont {S.}~\bibnamefont {Succi}}, \ and\ \bibinfo
  {author} {\bibfnamefont {H.~J.}\ \bibnamefont {Herrmann}},\ }\href {\doibase
  10.1103/PhysRevE.92.023309} {\bibfield  {journal} {\bibinfo  {journal} {Phys.
  Rev. E}\ }\textbf {\bibinfo {volume} {92}},\ \bibinfo {pages} {023309}
  (\bibinfo {year} {2015})}\BibitemShut {NoStop}%
\bibitem [{\citenamefont {Hoshen}\ and\ \citenamefont
  {Kopelman}(1976)}]{hoshen}%
  \BibitemOpen
  \bibfield  {author} {\bibinfo {author} {\bibfnamefont {J.}~\bibnamefont
  {Hoshen}}\ and\ \bibinfo {author} {\bibfnamefont {R.}~\bibnamefont
  {Kopelman}},\ }\href {\doibase 10.1103/PhysRevB.14.3438} {\bibfield
  {journal} {\bibinfo  {journal} {Phys. Rev. B}\ }\textbf {\bibinfo {volume}
  {14}},\ \bibinfo {pages} {3438} (\bibinfo {year} {1976})}\BibitemShut
  {NoStop}%
\end{thebibliography}%



\end{document}